\documentclass{ws-ijmpe}
\usepackage[super,compress]{cite}
\usepackage[breaklinks]{hyperref}
\hypersetup{colorlinks,urlcolor=black,citecolor=black,linkcolor=black,filecolor=black}
\usepackage{breakurl}
\usepackage{xcolor}
\usepackage{ulem}
\usepackage{orcidlink}
\usepackage{placeins}

\begin{document}

\markboth{Randy Dobler}{Didactic Bayesian Analysis for analysing Non-extensivity in Au-Au Collisions}

\catchline{}{}{}{}{}

\title{Bayesian Analysis of Non-extensive Parameters in Au-Au Collisions}

\author{
Randy Dobler\footnote{Frigadenstrasse 14, 8739 Rieden, Switzerland, randy.r.dobler@gmail.com}$^{1,2}$\orcidlink{0009-0006-0936-9623},
Juliana O. Costa$^{2}$\orcidlink{0000-0002-2824-4061},
Marcelo D. Alloy$^{2}$\orcidlink{0000-0001-9585-4188},
Débora P. Menezes$^{2}$\orcidlink{0000-0003-0730-6689}
}

\address{$^{1}$Department of Physics, ETH Zürich, Rämistrasse 101, Zürich, 8092, Switzerland}

\address{$^{2}$Departamento de Física, CFM, Universidade Federal de Santa Catarina, Rua Roberto Sampaio Gonzaga, s/n, Florianópolis, Santa Catarina 88.035-972, Brazil}

\maketitle

\begin{history}
\received{Day Month Year}
\revised{Day Month Year}
\end{history}

\begin{abstract}
In this work, a Bayesian statistical framework is employed to analyze particle yield ratios in Au-Au collisions, utilizing Non-Extensive Statistics (NES). Through Markov Chain Monte Carlo (MCMC) sampling, we systematically estimate key parameters, including the non-extensive factor \(q\), temperature \(T\), and chemical potential \(\mu.\) Our analysis confirms previous findings 
highlighting the suitability and robustness of Bayesian methods in describing heavy-ion collision data. A subsequent Bayes factor analysis does not provide definitive evidence favoring an Excluded Volume Hadron Resonance Gas (EV-HRG) model over the simpler NES approach. Overall, these results suggest that combining NES with Bayesian inference can effectively model particle distributions and improve parameter estimation accuracy, demonstrating the potential of this approach for future studies on relativistic nuclear interactions.
\end{abstract}


\ccode{PACS numbers:}


\section{Introduction}
Experiments involving collisions of heavy nuclei, such as Au-Au and Pb-Pb, in particle accelerators \cite{Achenbach:2023pba, PHENIX:2018lia, LUO201675} are crucial in investigating the fundamental structure of matter and the Quantum Chromodynamics (QCD) phase diagram \cite{chaudhuri2012short, LUO201675}. Conducted at RHIC (Relativistic Heavy Ion Collider) and the LHC (Large Hadron Collider), these experiments recreate extreme conditions of density and energy, analogous to those at the beginning of the universe.
During such collisions, intense exchanges of energy and momentum occur at relativistic speeds, compressing nuclear matter to extreme densities. As the nuclei collide, the strong nuclear interactions overcome nuclear binding, decomposing protons and neutrons into their fundamental constituents - quarks and gluons.  Measurements at RHIC and LHC suggest quark gluon plasma (QGP) temperatures at around 150 and 170 MeV, respectively \cite{Muller:2015jva, Shuryak:2008eq, lbl_qgp, cern_qgp}. These results indicate the initial temperature dependence on collision energy and confirm predictions from lattice QCD (LQCD) simulations \cite{Yasumichi_Aoki_2009}. The QGP behaves as a strongly interacting fluid, rapidly expanding and cooling. As the system cools, hadronization occurs, and quarks recombine into hadrons due to the fundamental properties of quarks and gluons, which, owing to color confinement, cannot exist as free particles under normal conditions.

In this work, we explore the hadronization process and its resulting particle production using statistical descriptions. Among the various modeling approaches available - such as thermal models - we adopt the statistical free gas model for its simplicity, broad applicability, and effectiveness in describing experimental data at chemical freeze-out \cite{menezes2007constraining, chiapparini2009hadron}.
For an even more detailed description of particle distributions, especially in systems that may exhibit fluctuations and correlations, NES can be integrated into free gas models as a generalization of the latter. Proposed by Constantino Tsallis \cite{Tsallis:1987eu}, NES extend the conventional Boltzmann-Gibbs statistics by introducing the non-extensive parameter $q$, which adjusts the degree of non-extensivity in the system. This approach enhances the applicability of free gas models without compromising their simplicity, allowing for the consideration of effects beyond strict thermodynamic equilibrium. NES allows for the treatment of more complex system characteristics, such as the ratios $\bar{p}/p$, $\bar{p}/\pi^{-}$, $\pi^{-}/\pi^{+}$, $K^{-}/K^{+}$, $K^{-}/\pi^{-}$, $\bar{\Lambda}^0/\Lambda^0$, and $\bar{\Xi}^{-}/\Xi^{-}$, which result from an Au-Au collision conducted at RHIC. This approach enables us to go beyond the extensivity and additivity assumptions of traditional statistics.
Studies such as \cite{COSTA2024138727, menezes2007constraining, chiapparini2009hadron, menezes2015non, Deppman:2019yno} demonstrate that NES is a valuable tool for analyzing particles formed during chemical freeze-out.

In this context, Bayesian inference emerges as a complementary tool to enhance the analysis of particle ratios. Based on Bayes' Theorem \cite{Gelman2013}, this approach enables the consistent adjustment of theoretical models to experimental data, integrating uncertainties and prior information. Studies such as \cite{rothkopf2019} and \cite{paquet2023} highlight the efficiency of Bayesian analysis in the context of heavy-ion collisions, while \cite{wesolowski2016} emphasizes its ability to ensure consistency even in complex parameter spaces. Thus, Bayesian inference, jointly with NES, provides a robust framework for describing and interpreting experimental data on particle multiplicities and ratios. However, traditional approaches to model validation often rely on subjective comparisons, lacking rigorous quantitative criteria to systematically assess the relative performance of competing statistical descriptions. To overcome this limitation, this paper leverages Bayesian model comparison via the Bayes factor, explicitly quantifying the relative plausibility of NES against the Excluded Volume Hadron Resonance Gas (EV-HRG) model. This statistical measure robustly incorporates both fit quality and model complexity, thereby offering a clear and objective means to guide the selection of theoretical models. The introduction of Bayesian comparative methods thus represents a methodological advance in the field, enhancing our ability to critically evaluate and interpret statistical models applied to heavy-ion collision data.

In this work, we first validate previously reported parameter estimates ($q$, $T$, and $\mu$) and systematically investigate the effectiveness and efficiency of Bayesian inference compared to traditional manual fitting techniques. By integrating NES with Bayesian inference, we provide an automated, reproducible, and statistically rigorous framework that reduces subjectivity and improves reliability in analyzing particle yield ratios. Additionally, we systematically explore which aspects of Bayesian analysis meaningfully impact the quality of results, identifying practical computational limits beyond which further refinement does not significantly enhance outcomes.

Subsequently, we perform the aforementioned Bayesian model comparison using Bayes factors to rigorously contrast the NES approach with the EV-HRG model, specifically examining whether the added complexity of excluded-volume effects is justified by improved descriptions of particle production data. This comparative analysis not only quantitatively evaluates the merit of additional model complexity but also provides insights into the thermodynamic conditions governing hadronization in relativistic nuclear collisions, thereby contributing both methodological innovation and deeper physical understanding.
\subsection{Bayesian Analysis Framework}
\label{steps}

To accurately estimate the non-extensive parameter $q$, the temperature $T$, and the chemical potential $\mu$, which are the relevant parameters in the analysis we intend to perform.
In the case investigated next, we employ a Bayesian statistical approach. This method provides a robust framework for quantifying uncertainties and extracting probabilistic information about model parameters by integrating theoretical predictions with experimental data.\\
We define the parameter space of interest as a three-dimensional vector:
\begin{equation}
\theta = (q, T, \mu).
\end{equation}
%
The Bayesian inference framework requires defining a prior probability distribution to encode previous knowledge about the parameters. A uniform prior distribution was chosen to reflect minimal prior assumptions and avoid introducing biases from previous analyses. This choice ensures that the parameter estimation is driven primarily by the experimental data rather than any specific preconceived distributions, allowing for an unbiased exploration of the parameter space.The uniform prior is defined as:
\begin{equation}
p(\theta) = \begin{cases}
\frac{1}{V_{\theta}}, & \text{if } \theta \text{ is within bounds,} \\[6pt]
0, & \text{otherwise,}
\end{cases}
\end{equation}

where $V_{\theta}$ represents the volume of the allowed parameter space.

The Bayesian analysis is performed by employing Markov Chain Monte Carlo (MCMC) methods to sample from the posterior distribution. This posterior distribution, which represents the probability of the parameters $\theta$ given the observed data, is defined by Bayes' theorem as:
\begin{equation}
p(\theta|\text{data}) \propto \mathcal{L}(\text{data}|\theta) \times p(\theta),
\label{eq1}
\end{equation}
where $\mathcal{L}(\text{data}|\theta)$ is the likelihood function and $p(\theta)$ is the prior probability distribution for the parameters.

The likelihood function, $\mathcal{L}(\text{data}|\theta)$, quantifies how well a given set of parameters $\theta$ describes the experimental data. In this context, the likelihood is inversely related to the reduced chi-square ($\chi^2/\text{dof}$), which is computed by comparing theoretical predictions to the measured particle ratios \cite{menezes2007constraining}. Therefore, the posterior probability distribution for the parameters is ultimately obtained by combining the prior information with this likelihood through the fundamental relationship of Bayes' theorem:
\begin{equation}
p(\theta|\text{data}) = \frac{\mathcal{L}(\text{data}|\theta) \times p(\theta)}{p(\text{data})},
\end{equation}
where $p(\text{data})$ is the evidence, a normalization constant that is often not explicitly calculated in MCMC methods as we are interested in the shape of the posterior. The proportionality in equation \ref{eq1} arises because $p(\text{data})$ is a constant for a given dataset.

In our numerical implementation, the sampling of the posterior distribution is performed via a Markov Chain Monte Carlo (MCMC) algorithm, specifically using the affine-invariant ensemble sampler. The computational implementation is structured as follows:

\begin{itemize}
    \item \textbf{Initialization}: Random sampling within the defined parameter bounds.
    \item \textbf{Burn-in phase}: A preliminary phase allowing the chains to equilibrate in the parameter space.
    \item \textbf{Sampling phase}: After discarding the initial (burn-in) samples, the parameter space is explored thoroughly to achieve robust statistical estimates.
\end{itemize}

To evaluate the convergence and efficiency of the sampling process, we consider the mean acceptance fraction of the proposed steps in the parameter space. Posterior distributions are visualized through corner plots, providing insights into parameter correlations and individual parameter uncertainties.

The best-fit values of the parameters are identified as those maximizing the posterior probability, i.e., minimizing the reduced $\chi^2$. These optimal parameters offer the best theoretical representation of the experimental data within the non-extensive statistical framework.
\subsection{Non-extensive statistics}
Rather than reiterating the well-established formalism of Boltzmann-Gibbs statistical mechanics, we proceed directly to its generalization via non-extensive statistics. Below, we outline the key definitions and mathematical structure of this extension.  Throughout this work, we use natural units, setting \( c = k_B = \hbar = 1 \).

 Non-extensive statistical mechanics, introduced by Constantino Tsallis in 1988, generalizes the classical Boltzmann-Gibbs framework by introducing a q-parameterized entropy ($S_q$) that accounts for non-additive systems. This formalism is particularly suited for describing complex systems characterized by long-range interactions, non-ergodic dynamics, multi-fractal geometries, or strong correlations that violate the standard assumptions of extensivity. \cite{Tsallis:1987eu, tsallis2021mecanica, Deppman:2019yno}. In contrast to the classical grand canonical approach - where systems are assumed to be extensive and additive - the non-extensive framework modifies the entropy and statistical weights, allowing a broader range of dynamical behaviors to be described. 

This generalization is based on the introduction of the \textit{q}-exponential function, which is the solution to the non-linear differential equation $dy/dx = y^q$. The \textit{q}-exponential function is defined as \cite{tsallis2009introduction}: 
\begin{align}
&e_q^{(+)} (x) = [1 + (q-1)x]^{\frac{1}{q-1}}, \quad x \geq 0, \label{regime1} \\
&e_q^{(-)} (x) = [1 + (1-q)x]^{\frac{1}{1-q}}, \quad x < 0, \label{qgener}
\end{align}
where, in analogy to the classical case, the argument $x$ is defined as $x = \beta(E_p \mp \mu N)$, with $\beta = 1/T$.
Here, $E_p = \sqrt{p^2 + m^2}$ denotes the energy of a particle with momentum $p$ and mass $m$, while $N$ represents the particle number.

The parameter $q$, known as the entropic index, characterizes the degree of non-extensivity of the system. For $q = 1$, Tsallis statistics reduces to the usual
statistics. The inverse function of the \textit{q}-exponential is the \textit{q}-logarithm, defined as \cite{tsallis2009introduction}: 
\begin{align}
&\ln_q^{(+)} (x) = \frac{x^{q-1} - 1}{q-1}, \quad x \geq 0, \label{regime2}, \\
&\ln_q^{(-)} (x) = \frac{x^{1-q} - 1}{1-q}, \quad x < 0. \label{refgime3}
\end{align}

The Tsallis entropy, a generalization of the classical Boltzmann-Gibbs entropy, is expressed as \cite{tsallis2011nonextensive}:
\begin{align}
    S_q = k_B \frac{1 - \sum_i p_i^q}{q-1}.
    \label{entropiadetsalliiiiis}
\end{align}
where \( p_i \) denotes the probability of the system being found in the microscopic state \( i \), satisfying the normalization condition
\begin{align}
\sum_i p_i = 1.
\end{align}

This entropy satisfies the pseudo-additivity property \cite{tsallis2009introduction}, reflecting its non-extensive nature:
\begin{align}
S_q{(A+B)} = S_q(A) + S_q(B) + (1-q) S_q(A) S_q(B),
\label{naoaditivatsallis}
\end{align}
which contrasts with the strict additivity of classical entropy. In the limit $q \to 1$, Tsallis entropy returns to the standard form, reinforcing its role as a consistent generalization.

The generalized partition function in the non-extensive formalism, crucial for determining Fermi-Dirac and Bose-Einstein distributions, is given by:
\begin{align}
    \log \mathcal{Z}_q = -\xi V \int \frac{d^3p}{(2\pi)^3} \sum_{r = \pm} \Theta(rx) \log_q^{(-r)} \left( \frac{e_q^{(r)}(x) - \xi}{e_q^{(r)}(x)} \right),
        \label{fungen365656}
\end{align}
where $\mathcal{Z}_q$ generalizes the classical grand partition function by incorporating the non-extensivity parameter $q$. The parameter $\xi$ distinguishes particle types, with $\xi = +1$ for bosons and $\xi = -1$ for fermions. The Heaviside function $\Theta(x)$ ensures the correct domain for the generalized logarithmic expressions.

In direct analogy to the classical approach - where the distribution function is derived from the derivative of $\ln \mathcal{Z}$ - the generalized particle distribution ($f_q$) in non-extensive statistics is:
\begin{align}
f_q = -\frac{1}{\beta} \frac{\partial \ln \mathcal{Z}_q}{\partial \varepsilon_k}.
\end{align}

From this, the generalized statistical distributions are obtained \cite{CONROY20104581}:
\begin{align}
&f_q^{(+)} (x) = \frac{1}{\left(e_q^{(+)}(x) - \xi \right)^q}, \quad x \geq 0, \\
&f_q^{(-)} (x) = \frac{1}{\left(e_q^{(-)}(x) - \xi \right)^{2-q}}, \quad x < 0.
\end{align}

Setting $\xi = -1$ for fermions and $\xi = +1$ for bosons yields the non-extensive counterparts of the Fermi-Dirac and Bose-Einstein distributions, respectively. These distributions preserve the formal structure of their classical counterparts but include the deformation parameter $q$, which modulates the deviation from standard equilibrium behavior. This demonstrates how non-extensive statistics extends the classical formalism to better capture systems exhibiting anomalous distributions or incomplete thermalization.

Analogous to the classical treatment - where the particle density is derived from integrating the occupation number over momentum space - the particle densities in the non-extensive framework are given by:
\begin{align}
\rho^{\text{Fermi Gas}{(+)}}_q &= \gamma \int \frac{d^{3}p}{(2 \pi)^3} f_q^{\text{FD}{(+)}}, \\
\rho^{\text{Fermi Gas}{(-)}}_q &= \gamma \int \frac{d^{3}p}{(2 \pi)^3} f_q^{\text{FD}{(-)}},
\end{align}
for fermions, and
\begin{align}
\rho^{\text{Bose Gas}{(\pm)}}_q = \gamma \int \frac{d^{3}p}{(2 \pi)^3} f_q^{\text{BE}{(\pm)}},
\end{align}
for bosons, where $\gamma = 2S + 1$ is the spin degeneracy, as in the classical case.

This formalism, therefore, enables the generalization of the free gas model to a non-extensive regime. It recovers the classical results in the limit $q \rightarrow 1$, but provides an enriched framework for studying hadronic systems in heavy-ion collisions where deviations from equilibrium may be significant.

\section{The Excluded Volume Model versus NES model}

In heavy-ion collision physics, accurately describing particle production requires models that realistically incorporate interactions and finite-size effects. While the NES incorporates a degree of non-ideality, it 
does not take into account 
effects due to the finite spatial extension of hadrons. The Excluded Volume (EV) model addresses these limitations by explicitly incorporating the finite spatial extension of hadrons, thereby modifying the thermodynamic equations and predictions for particle yields and densities.\\
The historical roots of the EV model trace back to pioneering work by Hagedorn and Rafelski \cite{Hagedorn1980} who first suggested finite-volume corrections to the ideal gas picture. Subsequent studies by Rischke, Gorenstein, and others refined this model, establishing its significance in understanding particle multiplicities and thermodynamic conditions at chemical freeze-out \cite{Rischke1991, Yen1998}. Recently, the EV model gained renewed attention due to its ability to describe hadron yields more accurately when compared to ideal gas approximations, particularly at high densities relevant for RHIC and LHC energies \cite{Andronic2018}.
While the NES model relies solely on non-extensive Tsallis statistics, characterized by the entropic parameter $q$, to capture non-equilibrium features such as long-range correlations and intrinsic temperature fluctuations, the EV model extends this framework by additionally incorporating hadronic volume exclusion effects. Thus, it modifies both the statistical distributions through $q$ and the thermodynamic treatment through excluded volume corrections.\\
The key theoretical difference lies in the treatment of interactions: the NES model alters the underlying momentum distribution to account for deviations from equilibrium, whereas the EV model, built on the same non-extensive formalism, further adjusts particle densities and pressures to reflect finite-size effects of hadrons. This combined approach leads to distinct predictions for particle yields, ratios, and thermodynamic quantities, especially in dense systems where spatial correlations become significant.
In contrast to the NES model, the EV model introduces the concept of eigenvolume for hadrons, effectively reducing the available phase space. For a system of hadrons, the pressure $P$ in the EV model is modified as:
\begin{equation}
P^{\text{EV}}(T, \mu) = \sum_i P_i^{\text{id}}(T, \tilde{\mu}_i),
\end{equation}
where $P_i^{\text{id}}$ is the ideal gas pressure of species $i$, and the modified chemical potential $\tilde{\mu}_i$ accounts for volume exclusion effects. For baryons and mesons, different eigenvolumes are often assumed ($v_{\text{bar}}$ and $v_{\text{mes}}$, respectively), leading to distinct volume corrections:
\begin{equation}
\tilde{\mu}_i = \mu_i - v_i P^{\text{EV}}(T, \mu),
\end{equation}
with $v_i = v_{\text{bar}}$ for baryons and $v_i = v_{\text{mes}}$ for mesons.
In the context of this work we analyze three different models. 
In statistical modeling of particle production, there are two primary approaches: analyzing absolute yields or particle fractions (ratios). Absolute yields provide direct insight into total particle production but can be heavily influenced by systematic uncertainties related to detector efficiencies, luminosities, and collision centralities. On the other hand, particle fractions, obtained by taking ratios of particle yields, mitigate many systematic uncertainties because they are less sensitive to experimental conditions and normalization effects.
In this work, particle fractions were ultimately chosen over absolute yields to ensure a fair and unbiased comparison between the EV and NES models. By using fractions, we minimize systematic uncertainties inherent in yield measurements, thereby focusing exclusively on the underlying physics described by the models. This choice facilitates clearer and more robust conclusions regarding model validity and performance in describing particle production.
Furthermore, in the excluded-volume formulation, one may choose either the hard-core radius \(R\) or the corresponding eigenvolume
\[
v = \frac{4\pi}{3}R^3,
\]
as the sampling parameter. Although these two parameterizations are mathematically equivalent, their induced priors differ significantly under uniform assumptions.

\subsection{Model selection}

This work reviews three distinct Python-based implementations of the EV-HRG model, developed and utilized for Bayesian analysis of experimental data and a subsequent Bayes factor comparison to the previous IHRG. A detailed description of these three models can be found in Randy Dobler's 2025 Master Thesis titled "Bayesian Analysis of Non-extensive Parameters in Au-Au Collisions" (available upon request). The boundary selection of the priors had to be adapted between the models to make sure the best-fit values were within those uniform priors. One of the explored models (A) gave slightly better results than a less complex version (B) whilst including considerably higher complexities. A third model (C) was, however, too simple to produce valuable predictions. Hence, we chose model B, providing an accurate model while keeping the complexity reasonable.

\subsubsection{Radius-Parameterized EV-HRG with q-Statistics}
\begin{figure}[htbp]
    \centering
    \includegraphics[width=1\linewidth]{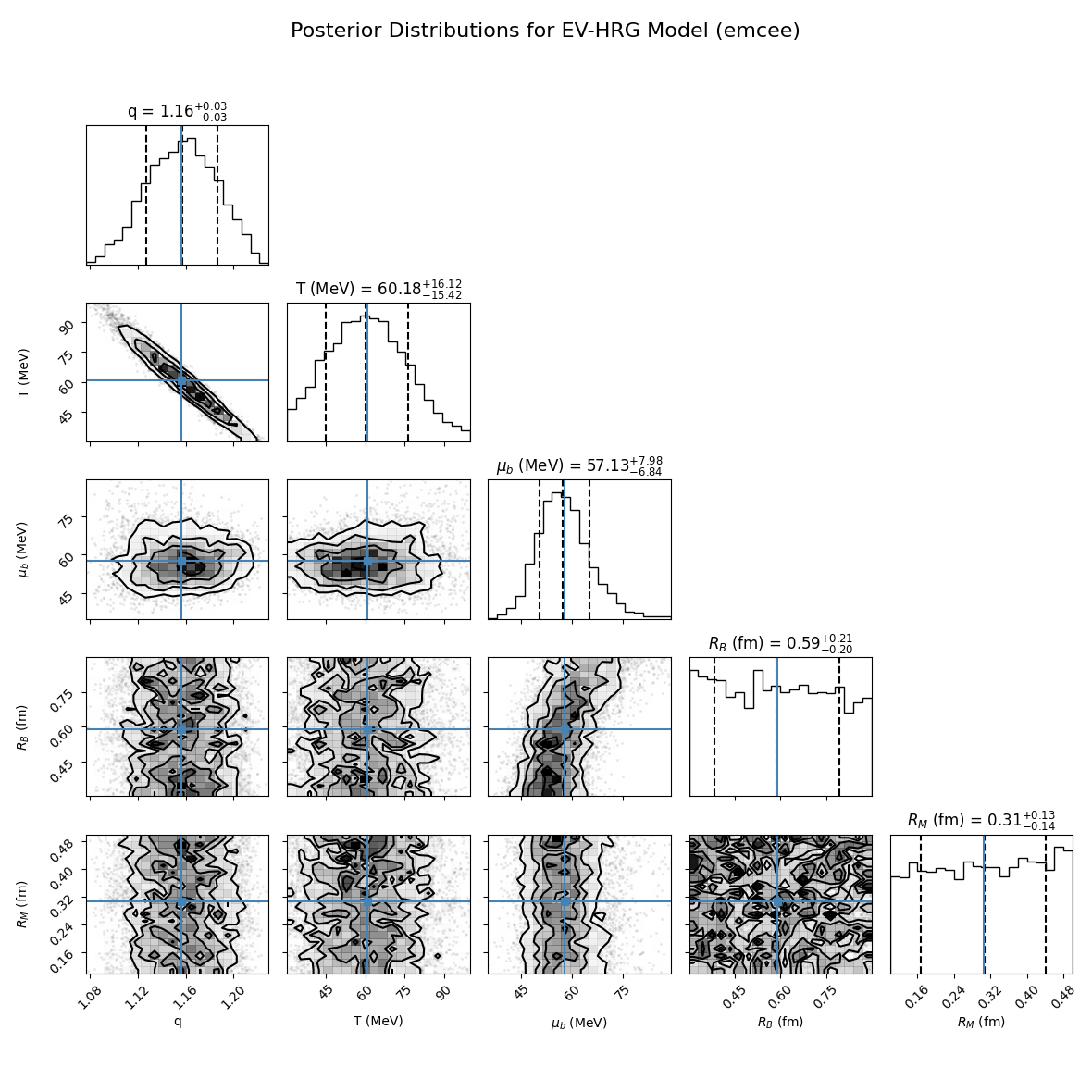}
    \caption{Posterior distributions for the five free parameters of the EV-HRG model with Tsallis statistics, obtained via MCMC sampling using the emcee sampler. The parameters include the non-extensivity index $q$, temperature $T$, chemical potential $\mu_B$, and the hard-core radii of baryons $R_B$ and mesons $R_M$. The contours represent the 1$\sigma$ and 2$\sigma$ credible regions, while the histograms along the diagonal show the marginalized distributions with median values and corresponding 68\% credible intervals.}
    \label{fig:bayes-factor3}
\end{figure}

Following, we describe the basic characteristics of the chosen model:
\begin{itemize}
    \item \textbf{Physical Model}: It incorporates an EV-HRG framework that allows for the inclusion of Tsallis non-extensive statistics, as usual characterized by  $q$. The excluded volume correction is applied by modifying chemical potentials as:
    \begin{equation}
        \mu_i^* = \mu_i - v_i P_{\text{EV}},
    \end{equation} where $P_{\text{EV}}$ is the self-consistently determined total excluded volume pressure,
    \begin{equation}
        P_{\text{EV}} = \sum_j P_j^{\text{id}}(T, \mu_j - v_j P_{\text{EV}}).
    \end{equation} Densities are then:
    \begin{equation}
      n_i^{\text{EV}} = n_i^{\text{id}}(T, \mu_i^*).  
    \end{equation}
    \item \textbf{Excluded Volume Parameterization}: A key feature is the use of separate excluded volumes for baryons ($v_B$) and mesons ($v_M$), which are derived from distinct hard-core radii ($R_B, R_M$) for these particle classes.
    \item \textbf{Statistical Treatment}: Particle distribution functions are generalized and can accommodate Tsallis statistics (if $q \neq 1$). The momentum-space integrals are performed using a custom Gaussian quadrature method.
    \item \textbf{Parameters}: Five primary free parameters are fitted: the Tsallis parameter ($q$), temperature ($T$), chemical potential ($\mu_B$), baryon radius ($R_B$), and meson radius ($R_M$).
    \item \textbf{Experimental Data}: The model is fitted against a set of experimental particle yield ratios, the same as in the original model.
    \item \textbf{Hadron Fraction List}: It employs a more extensive list of hadrons, more precisely hadron fractions. Specifically, the following:\\
    \(
    \overline{p}/p,\ \overline{p}/\pi^+,\ \pi^-/\pi^+,\ K^-/K^+,\ K^-/\pi^-,\ 
    \overline{\Lambda}/\Lambda,\ \overline{\Xi}/\Xi,\ K^{*0}/h^-,\ \overline{K}^{*0}/h^-
    \).
    \item \textbf{Conservation Laws}: The script explicitly solves for effective strangeness 
    and isospin-related 
    chemical potentials to satisfy constraints such as net strangeness and a specified isospin density (related to the system's charge-to-baryon ratio).
\end{itemize}
The results of this particular model can be seen in figure \ref{fig:bayes-factor3}.
It is worth noting that the convergence on the radii is rather poor. However, these are the best results that could be obtained after a significant computational effort.

\section{Bayes Factor and Model Comparison}

Beyond estimating parameters, Bayesian analysis offers a natural and rigorous framework for model comparison through the use of the \emph{Bayes factor} \( K \). Introduced and formalized by Harold Jeffreys \cite{Jeffreys:1998}, the Bayes factor quantifies the evidence provided by the data in favor of one model over another. This approach is particularly relevant when comparing competing physical models, such as different parameterizations or the inclusion versus exclusion of non-extensivity effects.\\
Given two models, \( \mathcal{M}_0 \) (simpler or null model) and \( \mathcal{M}_1 \) (alternative or extended model), the Bayes factor is defined as the ratio of their marginal likelihoods:
\begin{equation}
    K = \frac{P(D|\mathcal{M}_1)}{P(D|\mathcal{M}_0)} = \frac{\int P(D|\theta_1, \mathcal{M}_1)P(\theta_1|\mathcal{M}_1)\,\mathrm{d}\theta_1}{\int P(D|\theta_0, \mathcal{M}_0)P(\theta_0|\mathcal{M}_0)\,\mathrm{d}\theta_0},
\end{equation}
where \( P(D|\mathcal{M}_i) \) is the evidence or marginal likelihood of model \( \mathcal{M}_i \), obtained by integrating the likelihood over the prior distribution of the parameters \( \theta_i \). The Bayes factor thus penalizes model complexity automatically via the volume of the prior, embodying Occam's razor within the probabilistic structure of Bayesian reasoning \cite{mackay2003information}.\\
Occam's razor is a philosophical principle attributed to William of Ockham asserting that among competing hypotheses that explain the data equally well, the simplest one should be preferred. In the context of Bayesian model comparison, this principle is inherently encoded in the marginal likelihood \( P(D|\mathcal{M}_i) \), which averages the likelihood over the prior distribution of the model parameters \( \theta_i \). Models with more parameters typically have larger prior volumes, which dilute the average likelihood unless the data provide strong support. As a result, more complex models are automatically penalized unless justified by a significantly better fit to the data. The Bayes factor thus embodies Occam's razor within the probabilistic structure of Bayesian reasoning, favoring models that balance goodness-of-fit and simplicity.\\
Interpretation of the Bayes factor follows a graded scale, most notably summarized by Jeffreys in his benchmark classification, as shown in Table~\ref{tab:jeffreys-scale}. The logarithmic form \( \log K \) is often used for numerical stability and interpretability.
\begin{table}[h]
    \centering
    \caption{Interpretation of Bayes factor \( K \) according to Jeffreys' scale \cite{Jeffreys:1998}}
    \label{tab:jeffreys-scale}
    \begin{tabular}{c|l}
        \textbf{\( \log_{10} K \)} & \textbf{Strength of evidence for \( \mathcal{M}_1 \)} \\
        \hline
        \( < 0.5 \)   & Inconclusive \\
        \( 0.5 - 1 \) & Substantial \\
        \( 1 - 1.5 \) & Strong \\
        \( 1.5 - 2 \) & Very strong \\
        \( > 2 \)     & Decisive \\
    \end{tabular}
\end{table}

\section{Au-Au collision }

The Bayesian formalism described above has been applied to hadron production in Au-Au collisions and was subsequently compared to experimental data. 

For comparability reasons the conservation restraints on strangeness, number of baryons and electric charge have been kept the same as in Braun-Munzinger et al. 2001 \cite{BRAUNMUNZINGER200141}. They are given by the following expressions: 

\begin{equation}
V \sum_i n_i B_i = Q_B = Z + N,
\end{equation}
where \( V \) represents the volume in which the gas is confined, \( n_i \) denotes the number of particles of the \( i \)-th hadron species, and \( B_i \) is the baryon number. The quantity \( Q_B \) represents the total baryon charge of the gas. Additionally, \( Z \) stands for the total number of protons, while \( N \) denotes the total number of neutrons in the system. The value of the baryon number is calculated by considering the number of nuclei involved in the collision and the number of baryons in each nucleus. This analysis involves two gold (Au) nuclei. The most common isotope used in collisions is gold-197, which has a mass number ($A$) of 197. Thus, each Au nucleus has 79 protons ($Z$) and 118 neutrons ($N$), totaling 394 baryons in the system.

\begin{equation}
V \sum_i n_i S_i = Q_S = 0,
\end{equation}

where \( S_i \) denotes the strangeness of each gas particle species, and the total strangeness \( Q_S \) is zero.

\begin{equation}
V \sum_i n_i I_{3i} = Q_{I_3} = \frac{Z - N}{2},
\end{equation}
where \( I_{3i} \) represents the third component of the isospin associated with each particle, and \( Q_{I_3} \) is the total isospin. Thus, the total isospin is $Q_{I_{3}} = -19.5$ for each gold nucleus (or $-39$ for the entire system).
The $\chi^2$ is computed as follows:
\begin{equation}
\chi^2 = \sum_i \frac{(R_i^{\text{exp}} - R_i^{\text{theo}})^2}{\sigma_i^2},
\end{equation}
where $R_i^{\text{exp}}$ and $R_i^{\text{theo}}$ are experimental and theoretical particle ratios, respectively, and $\sigma_i$ is the uncertainty associated with the experimental measurements.
The parameter bounds for the Bayesian analysis are set as follows:
\begin{align}
1.0 \leq &\; q \leq 2.0,\\
30.0 \leq &T \leq 80.0~\text{MeV},\\
30.0\, \text{MeV} \leq &\mu \leq 80.0\, \text{MeV},
\end{align}

\begin{figure}[th]
\centerline{\includegraphics[width=8cm]{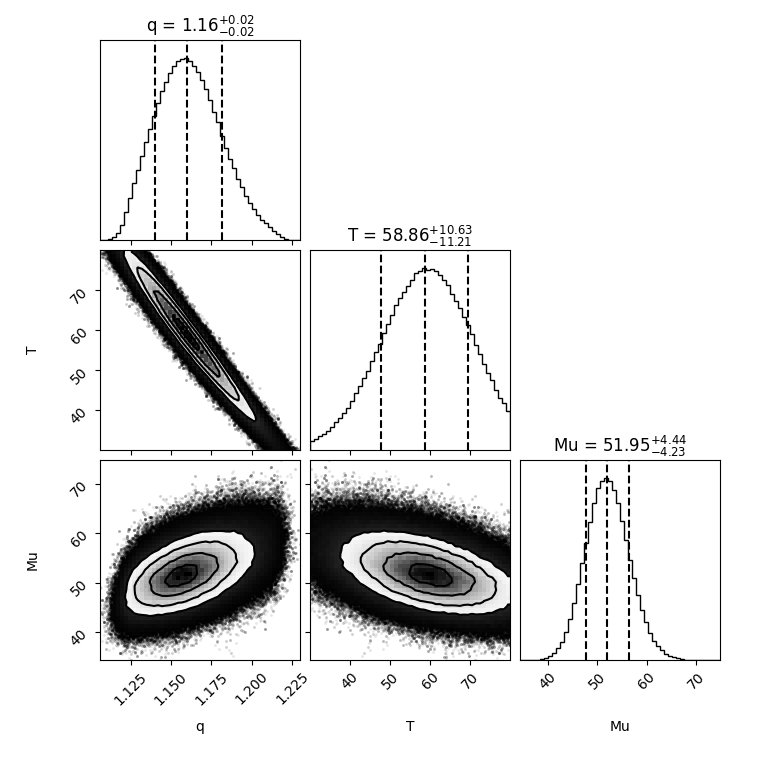}}
\caption{Corner plot for the Bayesian analysis framework analysing the parameter temperature $T$, chemical potential $\mu$ and the Tsallis parameter $q$. The number of walkers for the MCMC sampling was set at 400, the burn-in period at 200 and the number of steps at 5000. The results displayed reflect the highest likelihood point with the respective errors.}
\label{cornerplot}
\end{figure}

The corner plot resulting from the optimal Bayesian analysis is presented in figure \ref{cornerplot}. The corner plot displays the posterior probability distributions and pairwise correlations of the three key parameters. 
We next analyse the computational time and corner plot results for different choices of the three main steps described in \ref{steps}.

\begin{figure}[h]
    \centering
    \includegraphics[width=1\linewidth]{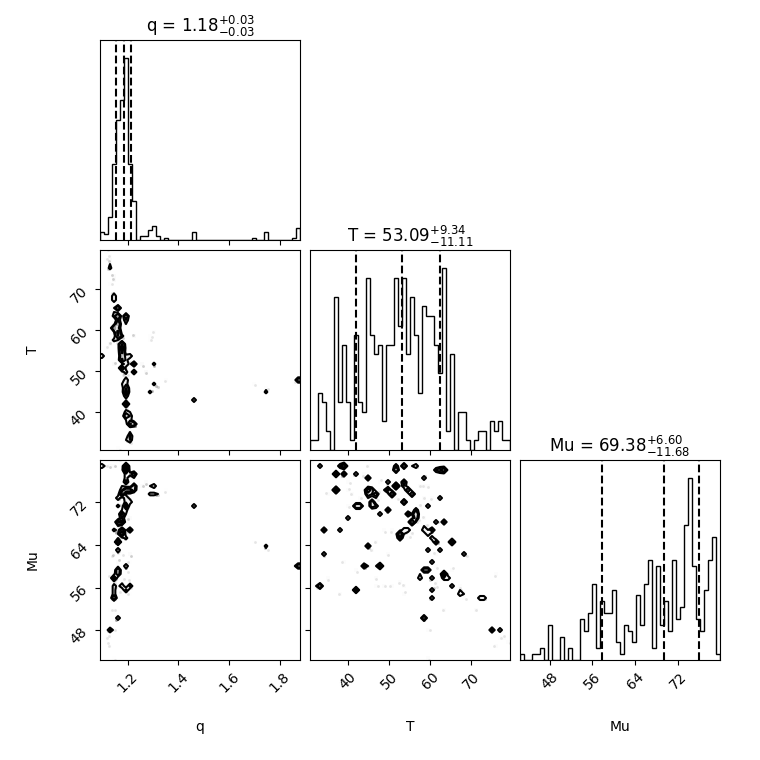}
    \caption{Corner plot for the Bayesian analysis framework analysing the parameter temperature $T$, chemical potential $\mu$ and the Tsallis parameter $q$. The number of walkers for the MCMC sampling was set at 6, the burn-in period at 6 and the number of steps at 60. The results displayed reflect the highest likelihood point with the respective errors.}
    \label{Figure:6}
\end{figure}

\begin{figure}[h]
    \centering
    \includegraphics[width=1\linewidth]{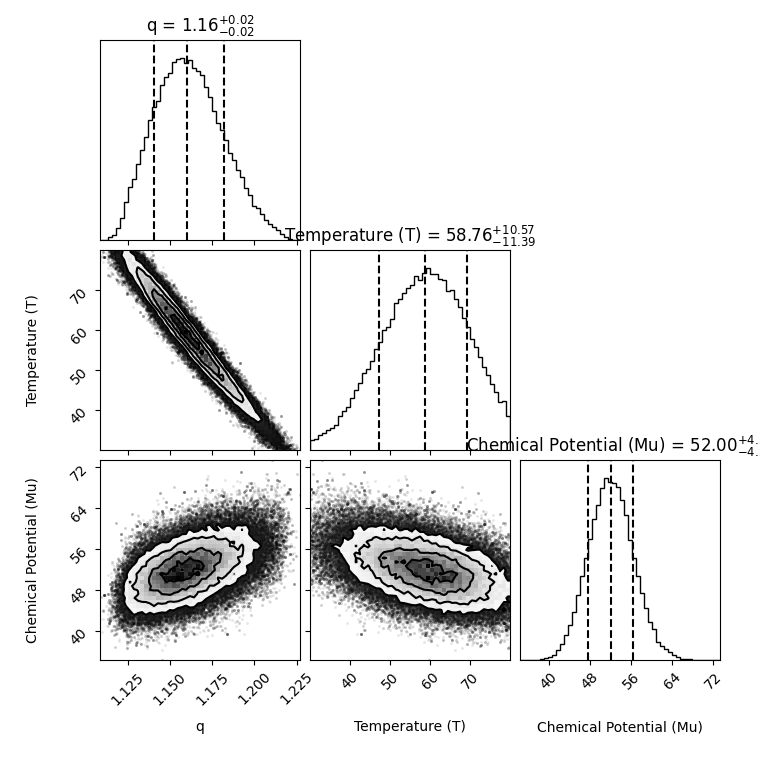}
    \caption{Corner plot for the Bayesian analysis framework analysing the parameter temperature $T$, chemical potential $\mu$ and the Tsallis parameter $q$. The number of walkers for the MCMC sampling was set at 200, the burn-in period at 200 and the number of steps at 1000. The results displayed reflect the highest likelihood point with the respective errors.}
    \label{Figure:200}
\end{figure}

\begin{figure}[h]
    \centering
    \includegraphics[width=1\linewidth]{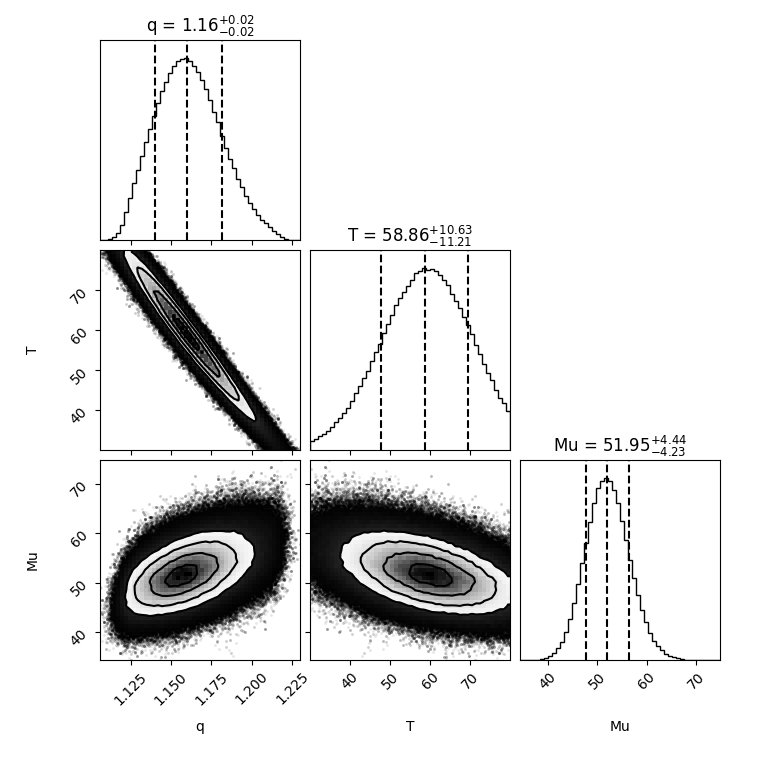}
    \caption{Corner plot for the Bayesian analysis framework analysing the parameter temperature $T$, chemical potential $\mu$ and the Tsallis parameter $q$. The number of walkers for the MCMC sampling was set at 300, the burn-in period at 200 and the number of steps at 5000. The results displayed reflect the highest likelihood point with the respective errors.}
    \label{Figure:400}
\end{figure}

The MCMC procedure comprised a preliminary burn-in phase of 400 steps per walker, followed by a production run of 200 steps per walker, totaling 5000 posterior samples. However, the result quality plateaued already at 200 walkers, a burn-in period of 200 and 1000 steps. This can be observed when comparing figures \ref{Figure:200} and \ref{cornerplot}. Although \ref{Figure:400} and \ref{cornerplot} consumed a lot more computing power in their calculation the results stayed the same. One could colloquially conclude that more is not better in this case. 
When comparing the results shown in Figures \ref{cornerplot} and \ref{Figure:6}, it becomes evident that even with minimal computational effort, one can obtain values that are consistently within the correct order of magnitude. However, the associated error margins often fail to encompass the true values. Furthermore, re-running the same code with identical parameters can sometimes yield poor results; results that may misleadingly appear to have converged, but toward an incorrect region of parameter space. This issue is difficult to detect without additional diagnostics. Yet, by examining the correlation plot, one can clearly determine whether the walkers were initialized in the correct region or if they started entirely off-target. If unsure about the results, a look at the correlation picture can tell if the results are trustworthy. In order to get consistently and reliably good results it is necessary in the case at hand to upon the number of walkers, steps, and the burn-in period from 6, 6, 60 to about 200, 200, 1000 respectively as seen in figure \ref{Figure:200}.
From these findings we could draw the following learnings:
\begin{itemize}
    \item There is an optimal number for these three parameters, at some point it doesn't get better anymore.
    \item There doesn't seem to be overfitting.
    \item Good quality results at low number of walkers, steps and burn in periods does not mean that these results can be reproduced with the same accuracy. A clear correlation picture indicates a good result within the actual error margins.
    \item Changing the visualization to more bins and away from grayscale can make life easier.
\end{itemize}
The optimal Bayesian analysis resulted in the median (50th percentile) values to be $q = 1.16^{+0.02}_{-0.02}$, $T = 58.86^{+10.63}_{-11.21}$ and $\mu = 51.95^{+4.44}_{-4.23}~\text{MeV} $, the highest likelihood point is at $q = 1.16$, $T = 58.48$ and $\mu = 51.67 $ compared to the previous values in ~\cite{costa2024} of $q = 1.16$, $T = 58.2$ and $\mu = 51.5 $. These values result in a $\chi^2$ of 10.569, 10.587 and 10.617 respectively. Divided by the degrees of freedom (dof) this results amount to 0.755, 0.756 and 0.758. The value of dof is 14, as there are 16 experimental data and 2 parameters
in the fit. The values for the different ratios are compared to experimental data from STAR, PHENIX, PHOBOS and BRAHMS and to the previous theoretical data from \refcite{costa2024} in Table \ref{Res}.

\begin{table}[pt]
\label{Res}
\tbl{Particle ratios of Au-Au collisions for $q = 1.16$, $T = 58.86$ and $\mu = 51.95$, 
     along the respective values of \cite{costa2024} in the Column named ``Costa 2024'' and experimental data at $\sqrt{s} = 130$\,GeV.}
{%
\begin{tabular}{@{}ccccc@{}}
\toprule
Ratio                & Bayesian & Costa 2024   & Exp. data ($\sqrt{s} = 130$\,GeV) & Exp. Ref. \\ 
\colrule
$\bar{p}/p$          & 0.6553   & 0.656767  & $0.65 \pm 0.07$                   & STAR\,\cite{ref2} \\
                    &          &           & $0.64 \pm 0.07$                   & PHENIX\,\cite{ref27} \\
                    &          &           & $0.60 \pm 0.07$                   & PHOBOS\,\cite{ref5} \\
                    &          &           & $0.64 \pm 0.07$                   & BRAHMS\,\cite{ref6} \\[4pt]

$\bar{p}/\pi^-$      & 0.0642   & 0.064067  & $0.08 \pm 0.01$                   & STAR\,\cite{ref20} \\[4pt]

$\pi^-/\pi^+$        & 1.0200   & 1.019848  & $0.95 \pm 0.06$                   & BRAHMS\,\cite{ref33} \\
                    &          &           & $1.00 \pm 0.02$                   & PHOBOS\,\cite{ref5} \\[4pt]

$K^-/K^+$            & 0.8280   & 0.829491  & $0.88 \pm 0.05$                   & STAR\,\cite{ref30} \\
                    &          &           & $0.78 \pm 0.13$                   & PHENIX\,\cite{ref28} \\
                    &          &           & $0.91 \pm 0.09$                   & PHOBOS\,\cite{ref5} \\
                    &          &           & $0.89 \pm 0.07$                   & BRAHMS\,\cite{ref33} \\[4pt]

$K^-/\pi^-$          & 0.1682   & 0.169266  & $0.149 \pm 0.02$                  & STAR\,\cite{ref30} \\[4pt]

$\Lambda^0/\bar{\Lambda}^0$ 
                    & 0.7696   & 0.770251  & $0.77 \pm 0.07$                   & STAR\,\cite{ref39} \\[4pt]

$\Xi^-/\bar{\Xi}^-$  & 0.8761   & 0.875823  & $0.82 \pm 0.08$                   & STAR\,\cite{ref19} \\[4pt]

$K^*/h^-$            & 0.0719   & 0.056983  & $0.06 \pm 0.017$                  & STAR\,\cite{ref39} \\[4pt]

$\bar{K}^*/h^-$      & 0.0626   & 0.062544  & $0.058 \pm 0.017$                 & STAR\,\cite{ref39} \\

\botrule
\end{tabular}}
\begin{tabnote}
Experimental data from STAR, PHENIX, PHOBOS, and BRAHMS collaborations.
\end{tabnote}
\end{table}
In figure \ref{cornerplot}, one can observe that the marginal distribution for $q$ is sharply peaked around $q = 1.16$, indicating that this parameter is well constrained by the data. This suggests a consistent degree of deviation from standard Boltzmann-Gibbs statistics across the sampled parameter space. The temperature $T$ and chemical potential $\mu$ are also reasonably well constrained, with highest likelihood points centered around $T \approx 58.5$~MeV and $\mu \approx 51.7$~MeV, respectively.

The joint distribution between $T$ and $\mu$ exhibits an elongated shape along a diagonal, which implies a degree of correlation between these two parameters.  
In contrast, $q$ shows minimal correlation with both $T$ and $\mu$, suggesting that its influence on the model fit is largely independent of the other parameters. The corner plot \ref{cornerplot} confirms that the Bayesian framework successfully constrains all three parameters and higlights interdependencies betweeen them.

One can hence observe that the Bayesian analysis confirmed the previous results on the Tsallis parameter $q$ and even allowed for a slight improvement in the values for $T$ and $\mu$. We did so by automating a process that has previously been done manually. The presented method can therefore be seen as a more efficient method for an effective minimization process.
\subsection{Bayes Factor Results} 
\begin{figure}[h]
    \centering
    \includegraphics[width=0.7\textwidth]{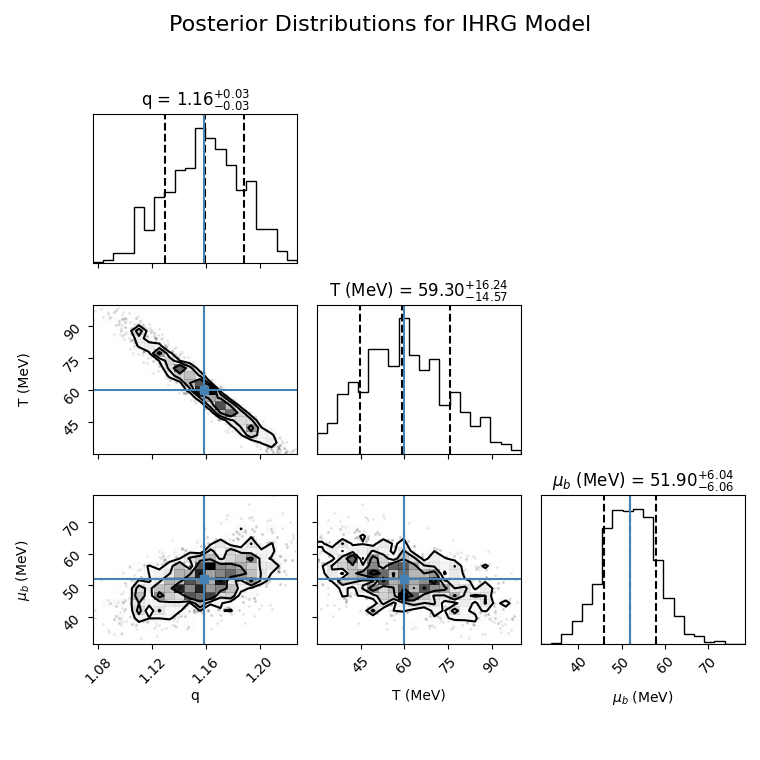}
    \caption{Corner plot for the Bayesian analysis of the original NES model.}
    \label{fig:bayes-factor}
\end{figure}

\begin{figure}[h]
    \centering
    \includegraphics[width=0.7\textwidth]{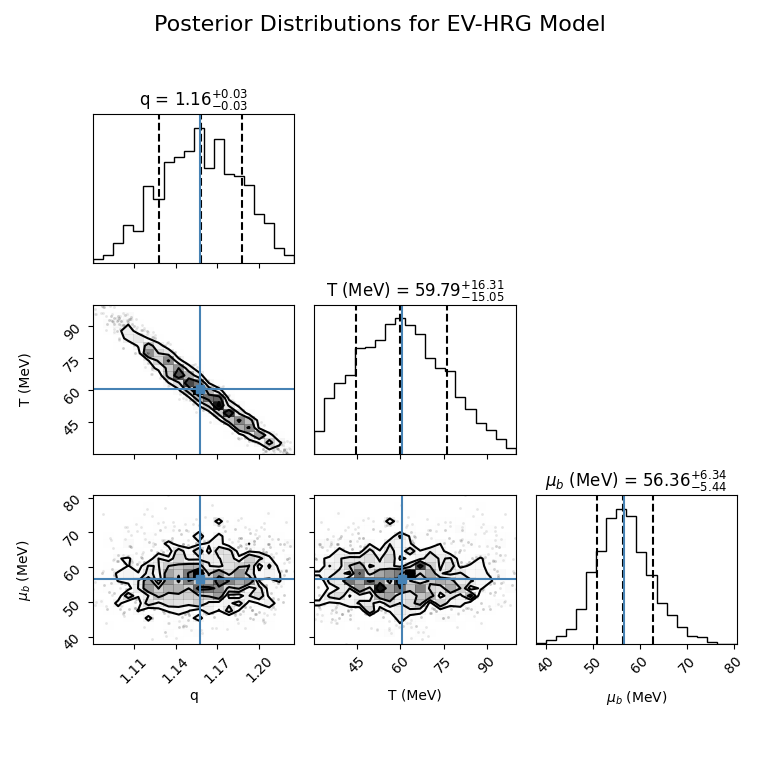}
    \caption{Corner plot for the Bayesian analysis of the final selected excluded volume model with parameters $R_B$=0.59 fm and $R_M$=0.31 fm}
    \label{fig:bayes-factor2}
\end{figure}

In a first step we could show that the NES model in Bayes factor comparison was heavily favoured over the ideal gas model with a K factor of 13.
Subsequently, with the analysis of the two models in place we proceeded to conduct a Bayes factor comparison between the EV and the NES (IHRG) model. In order to have a fair comparison with equal number of parameters we fixed in the EV model the exclude volume parameters to $R_B$=0.59~fm and $R_M$=0.31~fm respectively. We tested several nested sampling implementations for Bayesian model comparison. In particular, three ways of solving the pressure function were tested. Among the tested approaches, one provided consistently stable results and was adopted for the final Bayes factor evaluation. The outcome slightly favors the IHRG model over the EV-HRG model, with a Bayes factor of
\[
\log B = -0.268 \pm 0.290.
\]
However, according to Jeffrey's scale this factor has to be classified as inconclusive. Some runs with less extensive parameters on the nested sampling also showed a slight advantage for the excluded volume model. This inconclusive result suggests that, given the experimental uncertainties, the simpler NES model remains competitive against the more complex EV-HRG model. Therefore, unless substantially more precise data becomes available, incorporating volume exclusion effects may not yield clear improvements.\\
Despite this, it is important to note that the best-fit $\chi^2$ values for the EV-HRG model were in certain combinations lower than those of the IHRG, indicating a better point-wise fit. The preference for IHRG is thus driven not by accuracy but by Occam's razor embedded in the Bayesian framework: the increased complexity of the EV-HRG model is penalized in the integrated evidence. The final result from the Bayes factor analysis and the corner plots that were produced, can be seen in figures \ref{fig:bayes-factor} and \ref{fig:bayes-factor2}. One might say that the effects are minimal since our temperatures are very low, and increased temperature would make the volume corrections more prominent and hence relevant. However, within our framework, the $\chi^2$ difference at higher temperatures, in the case at hand 100 MeV, increased significantly in favor of the IHRG model with increasing temperature. The IHRG model gave a $\chi^2$ of 202, whereas the EV model had a $\chi^2$ of 457 with these ``ho'' parameters. One possible explanation might be that the eigenvolume parameters (or the radii from which they are derived) are not temperature-independent in our model. However, such dependencies might arise from medium modifications of hadron properties and more complex interaction models where the effective ``size'' parameter changes with thermodynamic conditions.
\section{Conclusions}
In this paper, we conducted a Bayesian analysis to estimate the non-extensive parameter \( q \), temperature \( T \), and chemical potential \( \mu \) in Au-Au heavy ion collisions, employing an NES framework. Our findings confirm previously obtained values for the parameter \( q \) while achieving slight improvements for \( T \) and \( \mu \). By integrating Bayesian inference with NES, we demonstrate a more efficient, statistically rigorous, and automated method compared to traditional manual fitting techniques, significantly reducing computational effort without compromising the quality of results.\\
The Bayesian approach introduced here effectively handles the inherent uncertainties in experimental data, providing clearer insights into parameter correlations and reducing subjectivity in the analysis process. This study underscores the utility of Bayesian methods in refining theoretical predictions of particle multiplicities and ratios in heavy ion collisions. Furthermore, the developed methodology can be extended to analyze other collision systems or alternative datasets, emphasizing its versatility and potential impact on future investigations of thermodynamic equilibrium and particle production under extreme conditions.

Beyond parameter estimation, this work also compared the NES with an excluded volume model through rigorous Bayesian parameters estimation. However, the Bayesian evidence did not support the inclusion of excluded volume corrections, especially in the low-temperature regime relevant for the analyzed collisions. The additional complexity introduced by eigenvolume-based models was not justified by the data, suggesting that such extensions must be treated with caution.

\section*{Acknowledgements}

This work is a part of the project INCT-FNA proc. No. 464898/2014-5. It is also supported by Conselho Nacional de Desenvolvimento Cient\'ifico e Tecnol\'ogico (CNPq) under Grants No. 303490/2021-7 (D.P.M.) and No 3247/2024 (J.O.C.) also thanking FAPESC for partial support.

\section{Appendix}
\FloatBarrier
\bibliographystyle{ws-ijmpe}
\bibliography{sample} 

\begin{thebibliography}{10}

\bibitem{Achenbach:2023pba}
P.~Achenbach {\em et~al.}, {\em Nucl. Phys. A} {\bf 1047}  (2024)   122874.

\bibitem{PHENIX:2018lia}
 PHENIX Collaboration (C.~Aidala {\em et~al.}), {\em Nature Phys.} {\bf 15}  (2019) 214, \href{http://arxiv.org/abs/1805.02973}{{\ttfamily arXiv:1805.02973 [nucl-ex]}}.

\bibitem{LUO201675}
X.~Luo, {\em Nuclear Physics A} {\bf 956}  (2016) 75, The XXV International Conference on Ultrarelativistic Nucleus-Nucleus Collisions: Quark Matter 2015.

\bibitem{chaudhuri2012short}
A.~K. Chaudhuri, A short course on relativistic heavy ion collisions  (2012).

\bibitem{Muller:2015jva}
B.~M\"uller, {A New Phase of Matter: Quark-Gluon Plasma Beyond the Hagedorn Critical Temperature}, in {\em {Melting Hadrons, Boiling Quarks - From Hagedorn Temperature to Ultra-Relativistic Heavy-Ion Collisions at CERN}: {With a Tribute to Rolf Hagedorn}\/},  ed. J.~Rafelski 2016, pp. 107--116.

\bibitem{Shuryak:2008eq}
E.~Shuryak, {\em Prog. Part. Nucl. Phys.} {\bf 62}  (2009) 48, \href{http://arxiv.org/abs/0807.3033}{{\ttfamily arXiv:0807.3033 [hep-ph]}}.

\bibitem{lbl_qgp}
{ Quark-gluon plasma}.

\bibitem{cern_qgp}
{ Quark-gluon plasma}.

\bibitem{Yasumichi_Aoki_2009}
Y.~Aoki, S.~Borsányi, S.~Dürr, Z.~Fodor, S.~D. Katz, S.~Krieg and K.~Szabo, {\em Journal of High Energy Physics} {\bf 2009} (Jun 2009)   088.

\bibitem{menezes2007constraining}
D.~Menezes, C.~Provid{\^e}ncia, M.~Chiapparini, M.~Bracco, A.~Delfino and M.~Malheiro, {\em Physical Review C—Nuclear Physics} {\bf 76}  (2007)   064902.

\bibitem{chiapparini2009hadron}
M.~Chiapparini, M.~Bracco, A.~Delfino, M.~Malheiro, D.~Menezes and C.~Provid{\^e}ncia, {\em Nuclear Physics A} {\bf 826}  (2009) 178.

\bibitem{Tsallis:1987eu}
C.~Tsallis, {\em J. Statist. Phys.} {\bf 52}  (1988) 479.

\bibitem{COSTA2024138727}
J.~O. Costa, I.~Aguiar, J.~L. Barauna, E.~Megías, A.~Deppman, T.~N. {da Silva} and D.~P. Menezes, {\em Physics Letters B} {\bf 854}  (2024)   138727.

\bibitem{menezes2015non}
D.~P. Menezes, A.~Deppman, E.~Meg{\i}as and L.~B. Castro, {\em The European Physical Journal A} {\bf 51}  (2015) 1.

\bibitem{Deppman:2019yno}
A.~Deppman, E.~Megias and D.~P. Menezes, {\em Phys. Rev. D} {\bf 101}  (2020)   034019, \href{http://arxiv.org/abs/1908.08799}{{\ttfamily arXiv:1908.08799 [hep-th]}}.

\bibitem{Gelman2013}
A.~Gelman, J.~B. Carlin, H.~S. Stern, D.~B. Dunson, A.~Vehtari and D.~B. Rubin, {\em Bayesian Data Analysis}, 3rd edn. (Chapman and Hall/CRC, Boca Raton, FL, 2013).

\bibitem{rothkopf2019}
A.~Rothkopf, {\em XIII Quark Confinement and the Hadron Spectrum}   (2019) \href{http://arxiv.org/abs/1903.02293}{{\ttfamily arXiv:1903.02293 [hep-ph]}}.

\bibitem{paquet2023}
J.-F. Paquet, {\em Review on model emulation and Bayesian methods}   (2023) \href{http://arxiv.org/abs/2310.17618}{{\ttfamily arXiv:2310.17618 [nucl-th]}}.

\bibitem{wesolowski2016}
S.~W. et~al., {\em EFT parameter estimation}   (2016) \href{http://arxiv.org/abs/1511.03618}{{\ttfamily arXiv:1511.03618 [nucl-th]}}.

\bibitem{tsallis2021mecanica}
C.~Tsallis, {\em Revista Brasileira de Ensino de F{\'\i}sica} {\bf 43}  (2021).

\bibitem{tsallis2009introduction}
C.~Tsallis, {\em Introduction to nonextensive statistical mechanics: approaching a complex world} (Springer, 2009).

\bibitem{tsallis2011nonextensive}
C.~Tsallis, { Nonextensive statistical mechanics: Applications to high energy physics}, in {\em EPJ Web of Conferences\/},  (2011), p. 05001.

\bibitem{CONROY20104581}
J.~Conroy, H.~Miller and A.~Plastino, {\em Physics Letters A} {\bf 374}  (2010) 4581.

\bibitem{Hagedorn1980}
R.~Hagedorn and J.~Rafelski, {\em Phys. Lett. B} {\bf 97}  (1980) 136.

\bibitem{Rischke1991}
D.~H. Rischke, M.~I. Gorenstein, H.~Stoecker and W.~Greiner, {\em Z. Phys. C} {\bf 51}  (1991) 485.

\bibitem{Yen1998}
G.~Yen, M.~Gorenstein, W.~Greiner and S.~Yang, {\em Physical Review C} {\bf 56}  (1997) 2210.

\bibitem{Andronic2018}
A.~Andronic, P.~Braun-Munzinger, K.~Redlich and J.~Stachel, {\em Nature} {\bf 561}  (2018) 321.

\bibitem{Jeffreys:1998}
H.~Jeffreys, {\em The Theory of Probability}, 3rd edn. (Oxford University Press, 1998).

\bibitem{mackay2003information}
D.~J.~C. MacKay, {\em Information Theory, Inference, and Learning Algorithms} (Cambridge University Press, 2003).

\bibitem{BRAUNMUNZINGER200141}
P.~Braun-Munzinger, D.~Magestro, K.~Redlich and J.~Stachel, {\em Physics Letters B} {\bf 518}  (2001) 41.

\bibitem{costa2024}
J.~O. Costa, I.~Aguiar, J.~L. Barauna, E.~Megías, A.~Deppman, T.~N. da~Silva and D.~P. Menezes, {\em Phys. Lett. B} {\bf 854}  (2024)   138727.

\bibitem{ref2}
e.~a. C.~Adler, {\em Phys. Rev. Lett.} {\bf 86}  (2001)   4778.

\bibitem{ref27}
F.~Messer, {\em Nucl. Phys. A} {\bf 698}  (2002)   511–514.

\bibitem{ref5}
e.~a. B.B.~Back, {\em Nucl. Phys. A} {\bf 721}  (2003)   C227–C230.

\bibitem{ref6}
B.~C. I.G.~Bearden, et~al., {\em Phys. Rev. Lett.}   (2001).

\bibitem{ref20}
e.~a. P.~Braun-Munzinger, {\em Phys. Lett. B} {\bf 518}  (2001)   41–46.

\bibitem{ref33}
B.~C. F.~Videbaek, et~al., {\em Nucl. Phys. A}   (2024) In press.

\bibitem{ref30}
H.~J. Specht, {\em Nucl. Phys. A} {\bf 698}  (2002)   341–359.

\bibitem{ref28}
H.~Ohnishi, {\em Nucl. Phys. A} {\bf 698}  (2002)   659–662.

\bibitem{ref39}
Z.~Xu, {\em Nucl. Phys. A}   (2024) In press.

\bibitem{ref19}
H.~Huang, { Proceedings of quark matter 2001}, in {\em Proceedings of Quark Matter 2001, Stony Brook, New York, Jan. 2001\/},  {\em Nucl. Phys. A}  (2024).
\newblock in press.

\end{thebibliography}
\end{document}